\documentclass{ws-p8-50x6-00}

\def\pmb#1{\setbox0=\hbox{#1}%
  \kern-.025em\copy0\kern-\wd0
  \kern.05em\copy0\kern-\wd0
  \kern-.025em\raise.0433em\box0 }
\def\bA{\pmb{$\cal A$}}

\def\gve{g_V^e}
\def\gae{g_A^e}
\def\rad2{\sqrt{2}}

\def\rumep2{\sqrt{1-\varepsilon^2}}

\def\tGE{\tilde{G}_E}
\def\tGM{\tilde{G}_M}
\def\tGA{\tilde{G}_A}
\def\gaesse{g_A^{(s)}}
\def\mus{\mu_s}
\def\rdues{r^2_s}
\def\Ax{ {\cal A}_x }
\def\Az{ {\cal A}_z }
\def\Ae{ {\cal A}_{LR} }
\def\tep{ \vartheta_{e'} }  
\def\t2g{\tan^2(\vartheta_{e'}/2)}
\begin{document}
\title{Hadronic degrees of freedom}

\author{Marco Traini}
 
\address{Dipartimento di Fisica \\
Universit\`a degli Studi di Trento \\
I-38050 POVO (Trento), Italy\\
{\sl and} \\
INFN G.C. Trento\\
E-mail: traini@science.unitn.it}

%%%%%%%%%%%%%%%%%%%%%%%%%%%%%%%%%%%%%%%%%%%%%%%%%%%%%%%%%%%%%%
% You may repeat \author \address as often as necessary      %
%%%%%%%%%%%%%%%%%%%%%%%%%%%%%%%%%%%%%%%%%%%%%%%%%%%%%%%%%%%%%%

\maketitle

\abstracts{I report on the research activities performed under the (italian)
MURST-PRIN project "{\sl Fisica Teorica del Nucleo e dei sistemi a pi\'u corpi"}
covering part of the topics on hadronic degrees of freedom. The most recent
achievements in the field are summarized focusing on the specific role of the nuclear 
physics community.
}
 
\section{Introduction}
\label{Intro}

The point of view I take to summarize some of the research activities
in Italy in the field of hadronic physics, is determined by the (today accepted)
fundamental degrees of freedom of the strong interactions, namely quarks and gluons.
The Sec. \ref{effquarks} is devoted to the link of the fundamental theory of strong
interactions, Quantum Chromodynamics (QCD), with effective degrees of freedom
often used to study specific problems, in particular those degrees of freedom that
played a peculiar and historical role in the investigation of hadronic systems:
the constituent quarks and the related quark models. 
In Secs. \ref{emNRQM} and \ref{emLFQM} I discuss recent results on the
electromagnetic interactions with hadrons at low energy studied within non-relativistic
and relativistic quark models, respectively. Sec. \ref{EW} is devoted to the
discussion of the electroweak structure of the nucleon and Sec. \ref{DIS} to deep 
inelastic scattering. Conclusive remarks are drawn in Sec. \ref{CONCL}.

\section{QCD and effective degrees of freedom}
\label{effquarks}

Why does the non-relativistic quark model (NRQM) reproduces {\it at the quantitative}
level several hadron properties? And why - by an appropriate choice of the
parameters - also several other models, totally different from NRQM ({\it
e.g.}, the chiral bag model) often fit the data? In a series of papers Morpurgo 
and Dillon\cite{Moa,DiMo} tried to answer to the previous question by using an original 
approach called {\it General parametrization method}. Following the Morpurgo's
review paper of ref.\cite{Moa}, I summarize the approach in the case example
of the baryon masses.

\subsection{General parametrization: basic ideas}

The QCD mass $M_{\rm B}$ of the baryon B is the expectation value of the exact QCD
Hamiltonian, $H_{\rm QCD}$ on the exact lowest state $\Psi_{\rm B}$ of B:
\begin{equation}
M_{\rm B} = \langle \Psi_{\rm B} |H_{\rm QCD}|\Psi_{\rm B}\rangle\;.
\label{QCDmassB}
\end{equation}
One can write
\begin{equation}
|\Psi_{\rm B}\rangle = {\cal V}\,|\phi_{\rm B}\rangle\;,
\label{Vop}
\end{equation}
where the auxiliary state $|\phi_{\rm B}\rangle$ is a three-quark-no-antiquark-no-gluon 
state in Fock space and it reduces to the simplest NRQM.

Whereas $|\phi_{\rm B}\rangle$ is quite simple, ${\cal V}$ is exceedingly complicated. Indeed
${\cal V}$ must take into account: 

\noindent i) the dress of the auxiliary state $|\phi_{\rm B}\rangle$
with $q\bar q$ pairs and gluons in order to transform it into the exact state which
assumes the form:
\begin{equation}
|\Psi_{\rm B}\rangle = |qqq\rangle + |qqq,\bar q q\rangle + |qqq,Gluons\rangle + ...\;;
\label{exactwf}
\end{equation}

\noindent ii) the mixing of $SU_3$-flavor configurations;

\noindent iii) the Foldy-Wouthuysen transformation of the quark spin states from static 
4-spinors (with upper components $(0,1)$ or $(1,0)$)) to Dirac 4-spinors.
\subsection{Baryon masses}
By using the ${\cal V}$ transformation, Eq.(\ref{QCDmassB}) becomes
\begin{equation}
M_{\rm B} = \langle \Psi_{\rm B} |H_{\rm QCD}|\Psi_{\rm B}\rangle =
\langle \phi_{\rm B} |{\cal V}^\dag\, H_{\rm QCD}\,{\cal V}|\phi_{\rm B}\rangle\;.
\label{paramassB}
\end{equation}

As final result one gets an expression for the "parametrized mass" operator 
which depends on the flavor and spin operators of the three quarks, namely:
\begin{eqnarray}
\hat {\cal M} = &M_0& + B \sum_{i} P_i^S + C \sum_{i>k} \left({\vec \sigma}_i\cdot 
{\vec \sigma}_k\right) + D \sum_{i>k} \left({\vec \sigma}_i \cdot {\vec \sigma}_k \right)
\left(P_i^S+P_k^S\right) + \nonumber \\
&+& E \sum_{i\neq k \neq j} \left({\vec \sigma}_i\cdot {\vec \sigma}_k\right)\,P_j^S +
a\sum_{i>k} P_i^S P_k^S + b \sum_{i>k} \left({\vec \sigma}_i \cdot {\vec \sigma}_k \right)
\left(P_i^S P_k^S\right) + \nonumber \\
&+& c \sum_{i \neq k \neq j} \left({\vec \sigma}_i \cdot {\vec \sigma}_k \right)
\left(P_i^S+P_k^S\right) P_j^S + d\; P_1^S P_2^S P_3^S\;,
\label{PMoperator}
\end{eqnarray}
where $P_i^S$ are projection operators selecting the strange $S$-quark, and 
${\vec \sigma}$ the quark spin Pauli matrices; $M_0$, $B$, $C$,...$d$ are parameters and 
only the combination $a+b$ is relevant for the masses. Few comments on 
the general parametrization (\ref{PMoperator}) are in order:

\noindent i) it has been obtained for the baryon 
masses but can be extended to many physical quantities of the lowest multiplets 
of hadrons (both baryons and mesons), like magnetic moments, semileptonic matrix elements
etc..

\noindent ii) it shows a large similarity with the NRQM result, but it is exact 
within QCD and fully relativistic (even if non-covariant because derived in the rest frame 
of the nucleon). In particular it takes into accont the flavor breaking term 
$\Delta m \bar \Psi P^S \Psi$ in the QCD Lagrangian to all orders.

\noindent In order to better understand the approach, let me introduce
a provocative question: the masses of the lowest octect and decuplet 
baryons are eight ($N$, $\Lambda$, $\Sigma$ , $\Xi$, $\Delta(1232)$, $\Sigma(1385)$, 
$\Xi(1530)$ and $\Omega$) and Eq.(\ref{PMoperator}) contains eight parameters ($M_0$, $B$, 
$C$, $D$, $E$, $a+b$, c, d): why is it convenient to deal with the eight parameters rather 
than with the eight masses directly?

The key point is that the dynamical consequence of QCD produces a hierarchy of the 
paramenters which depends on the number of the indices in Eq.(\ref{PMoperator}). 
Some of them result to be larger than others,
a peculiarity absent in the conventional $SU_3$ group-theoretical parametrization of the 
flavor-breaking terms. Specifically one gets (the values are in MeV):

\noindent $M_0 = 1076$, $B=192$, $C=45.6$, $D=-13.8\pm0.3$, $(a+b)= -16.0\pm 1.4$, 
$E=5.1\pm 0.3$, $c=1.1\pm 0.7$, $d=4\pm 3$.

\noindent Omitting all terms beyond $E$ in Eq.(\ref{PMoperator}) (that is keeping first order
flavor-breaking terms only) one obtains the Gell-Mann-Okubo mass formula.

The coefficients decrease for terms with increasing number of indices and, at equal number 
of indices, they decrease increasing the order in flavor breaking (that is the number of 
$P^S$ factors). Each additional $P^S$ factor implies a reduction of $\approx 0.3$
(e.g. the ratio $D/C$). For each additional index (at equal number of $P^S$ factors) the
reduction is $\approx 0.37$ (e.g. the ratio $E/D$).

An example is given by the Coleman-Galshow (CG) mass formula
\begin{equation}
p-n = \Sigma^+ -\Sigma^- +\Xi^- - \Xi^0\;,
\label{CGmf}
\end{equation}
derived assuming unbroken $SU_3$ flavor. Since flavor is violated (in the baryon octect)
by $\approx 33$\%, a similar violation of CG is expected.
However by using the new recent measured vale of $\Xi^0$, $1314\pm 0.06\pm 0.2$ MeV,
one gets, for the left and right side of Eq. (\ref{CGmf}), the impressive values
$
{\rm l.h.s.} = -1.29\;{\rm MeV}\;;
$
$
{\rm r.h.s.} = -1.58  \pm 0.25\;{\rm MeV}\,.
$

The general parametrization approach is able to solve the mistery of the spectacular
precision of the CG mass formula\cite{DiMo}. As matter of fact the expression (\ref{CGmf}) 
is valid to all orders in flavor breaking terms with the omission of the terms 
with 3-quarks indices. The hierarchy in the general parametrization expansion allows
an estimate of the relative contribution of the three-quark terms: 
$(1/3)^3 \approx 4 \cdot 10^{-2}$ to be compared with the level of precision of the
CG mass formula $((1.58-1.29) \pm 0.25)/8 \approx (4 \pm 3)\%$, since the mass 
difference $\Sigma^- - \Sigma^+ \approx 8$ MeV. The present case is a peculiar case
where, thanks to the hierarchy in the parametrization, an estimate of an effect due 
to strong interaction can be given and found tiny as expected.

\section{NRQM: electromagnetic transitions}
\label{emNRQM}

A consequence of the large success of the Constituent Quark Models (CQM) is
the proliferation of a large class of different approaches. 

In particular various models have been proposed for the internal 
baryon structure. A quite common feature of many models
is that, despite of the use of different ingredients, they 
are able to give a satisfactory description of the baryon spectrum and, in 
general, of the nucleon static properties (cfr. the discussion of the 
previous section).

The obvious reason is that the study of hadron spectroscopy is not 
sufficient to distinguish among the various forms of quark dynamics
and other observables, such as the electromagnetic transition form 
factors and the strong decay amplitudes, are 
important in testing models for the internal structure of the hadrons. 
As an example I discuss a recent calculation\cite{aiello98}
of the electromagnetic transition form factors using different potential models
within a specific approach which includes three-body forces within 
an hypercentral approximation.

\subsection{Transition form factors and the hypercentral potential model}

The electromagnetic transition form factors, 
$A_{1/2}(Q^2)$ and $A_{3/2}(Q^2)$, are defined as the transition matrix elements of the 
transverse electromagnetic interaction, $H_{e.m.}^t$, between the nucleon, 
$N$, and the resonance, $B$, states:
\begin{equation}
A_{J'_z}(Q^2)= \langle B, J', J'_z | H^t_{em}| N, 
J={1}/{2}, J_{z}={1}/{2}\rangle
\end{equation}

\noindent Following Aiello {\it et al.}, I compare various models:

\noindent 1) a potential which retains only the hypercoulomb 
and the linear confinement terms 
fitted to the spectrum, plus a standard hyperfine interaction;

\noindent 2) an analytical model which corresponds to the previous 
approximation 
plus a hyperfine interaction with a sooth $x$-dependence;

\noindent 3) a potential which has the property of 
reproducing exactly the dipole fit of the proton form factor;

\noindent 4) the harmonic oscillator with the parameter  $\alpha=0.229$ GeV  
which reproduces the proton charge radius;

\noindent 5) the harmonic oscillator  with the parameter $\alpha=0.410$ GeV 
corresponding to a confinement radius of the order of 
$0.5$ fm, required in order to reproduce the
$A^{p}_{3/2}$ at $Q^2=0$ for  the $D_{13}(1520)$-resonance. 

\begin{figure}[t]
\begin{center}
\mbox{\psfig{file=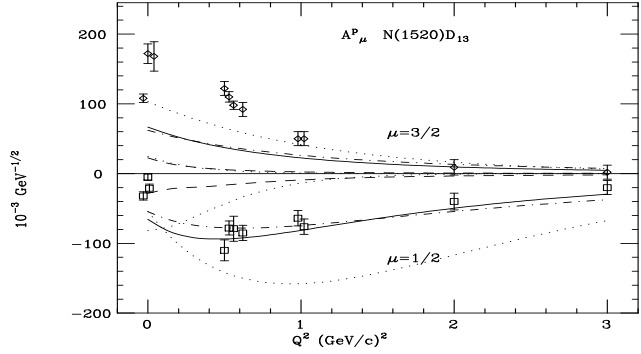,width=0.50\linewidth,height=0.60\textheight, angle=90}}
\end{center}
%fig 1.
\caption{Comparison between the experimental data for the transition form 
factors $A^p_{3/2}$,$A^p_{1/2}$ for the $D_{13}(1520)$-resonance and the 
calculations with the potentials 1) 
(full curve), 2) (dot-dashed curve), 3) (dashed curve), 4) (the dotted 
curve with the stronger damping) and 5) (the dotted curve with the softer 
damping).}
\label{tffHyp1}
\end{figure}
\begin{figure}
\begin{center}
\mbox{\psfig{file=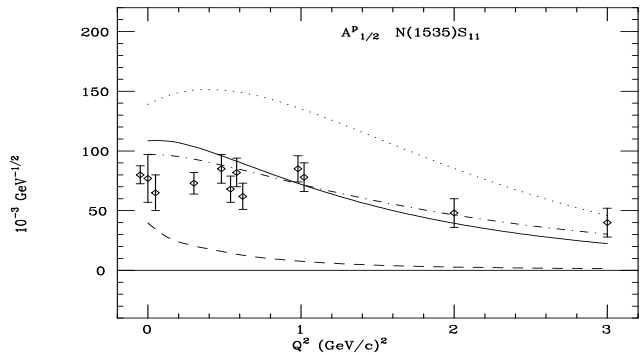,width=0.50\linewidth,height=0.60\textheight, angle=90}}
\end{center}
%fig 1.
\caption{As in previous Figure, for the $A^p_{1/2}$ of 
the $S_{11}(1535)$-resonance, calculated with the 
potentials 1), 2), 3) and 5).}
\label{tffHyp2}
\end{figure}

In Fig.~\ref{tffHyp1} the proton helicity amplitudes for 
$D_{13}(1520)$-resonance.  The  potentials 1) and 2) give rise 
to similar results.  They both fit the energy levels and lead to a 
confinement radius of the order of $0.5$ fm. 
The medium $Q^2$-behaviour is good but they fail to reproduce well the 
data at low $Q^2$ especially in the $A^p_{{3}/{2}}$ case.

The results are however very different from potential 1) and 2)
and in the $A^p_{{1}/{2}}$ case also far from the data. The 
potential which reproduces exactly the dipole form factor, 3), gives too 
damped results; the same happens for the h.o. with the correct proton radius, 
which causes a too strong damping in the wave functions. 

Similar conclusions can be drawn from Fig.~\ref{tffHyp2}, where the results for 
the $S_{11}(1535)$-resonance are shown. One can see that reproducing the 
elastic form factor is not a guaranty for describing also the transition 
form factors. 
None of the CQM considered can explain adequately the transition form 
factors at low momentum transfer. The discrepancy
indicates that some important effect at low momentum transfer is missing,
like polarization effect of the Dirac sea, not included in  CQMs. 
The calculations, at variance with what expected, are in agreement with 
the few existing data at $Q^2=1-2$ (GeV/c)$^2$, that is outside the range 
of applicability of a non relativistic description.

The problem of a relativistic description is still open\cite{desanctis98}  
and I discuss relativistic extensions in Sec.\ref{emLFQM}.

\subsection{Inelastic photon scattering and the magnetic moment of 
the $\Delta$(1232)}

As further example of CQM application let me briefly discuss inelastic photon scattering
and its relation with the magnetic moment of the $\Delta$(33) resonance as
investigated by Drechsel {\it et al.} in ref.\cite{drechsel00}.

The static properties of baryons are an important testing ground for
QCD based calculations in the confinement region. However, little 
experimental information is 
available for hadrons outside of the ground state SU(3) octet.   
In view of the short life-time of the resonances, such
information has to come from a detailed analysis of intermediate
states. As a result of many experimental and theoretical efforts, the
Particle Data Group (1998) quotes a value of
$\mu_{\Delta^{++}}= (5.6\pm 1.9)\mu_N$ for the magnetic
dipole moment of the $\Delta^{++}$ resonance. The large error bar is
due to large nonresonant processes, external bremsstrahlung by
initial and final state particles, and a strong background due to 
interactions in both the initial and final states.
A much cleaner experiment would be an
electromagnetic excitation of the nucleon leading to the $\Delta$
resonance with subsequent emission of a real photon followed by the
decay into a nucleon and a pion. The process $\gamma+p\rightarrow
\gamma'+p'+\pi^0$ would be particularly favorable, because the signal
is less disturbed by the external bremsstrahlung background.

Unfortunately, the A2 collaboration at MAMI working with the TAPS
detector has only been able to see 3-photon events
at a rate corresponding to a cross section of tens of nanobarns.
Drechsel {\it et al.} point out that one only expect
total cross sections for this process in the range of 5-10 nb, which
is probably at the limit of the present experimental accuracy. For
this purpose they have calculated both elastic (Compton) and inelastic
photon scattering through the $\Delta$ resonance in the constituent
quark model. The expectations based on such a simple model show that 
the integrated cross section for this process
should indeed be very small, namely of the order of 5~nb. 

The electromagnetic moments of baryon resonances are
among the most evasive properties of hadrons. The extremely weak
signals for these moments are at the very limits of even the most
advanced experimental techniques. However, such data would be
invaluable for our understanding of QCD in the confinement region, and
dedicated experiments are certainly desirable.

\section{Covariant Quark Models: electromagnetic form factors}
\label{emLFQM}

In the present section I discuss the relativistic extension of CQM.
In particular a relativistic light-front (LF) constituent-quark 
(CQ) model and the investigation of transition
electromagnetic hadron form factors~\cite{roma1,roma2} in the 
momentum transfer region relevant for the experimental research programme at
TJNAF. The main features of the model are: i) 
eigenstates of a mass operator which reproduces a large part of the hadron
spectrum; ii) a one-body current operator with phenomenological Dirac 
and Pauli form factors for the CQ's.
 The CQ's are assumed to interact via the $q - q$ potential of
Capstick and Isgur (CI), which includes a linear confining term 
and an effective one-gluon-exchange (OGE) term. 
The latter produces a huge amount of high-momentum components in the baryon wave
functions and contains a central Coulomb-like potential, a
spin-dependent part, responsible for the hyperfine splitting of baryon masses,
and a tensor part. A comparable amount of high momentum components was obtained
with the $q - q$ interaction based on the exchange of the
pseudoscalar Goldstone-bosons. This fact suggests that the hadron
spectrum itself dictates the high momentum behaviour in hadron wave functions
as furtherly suggested by the recent calculations of electromagnetic static
form factors within the Goldstone-bosons potential model\cite{paviaff}.
In the following I review results for the 
transition form factors for $J \leq {3}/{2}$ hadrons.

\begin{figure}[t]
\begin{center}
%\figurebox{30pc}{60pc}{} % to have a box alone
\epsfxsize=30pc % will enlarge or reduce the postscript figures based on the xsize
\epsfbox{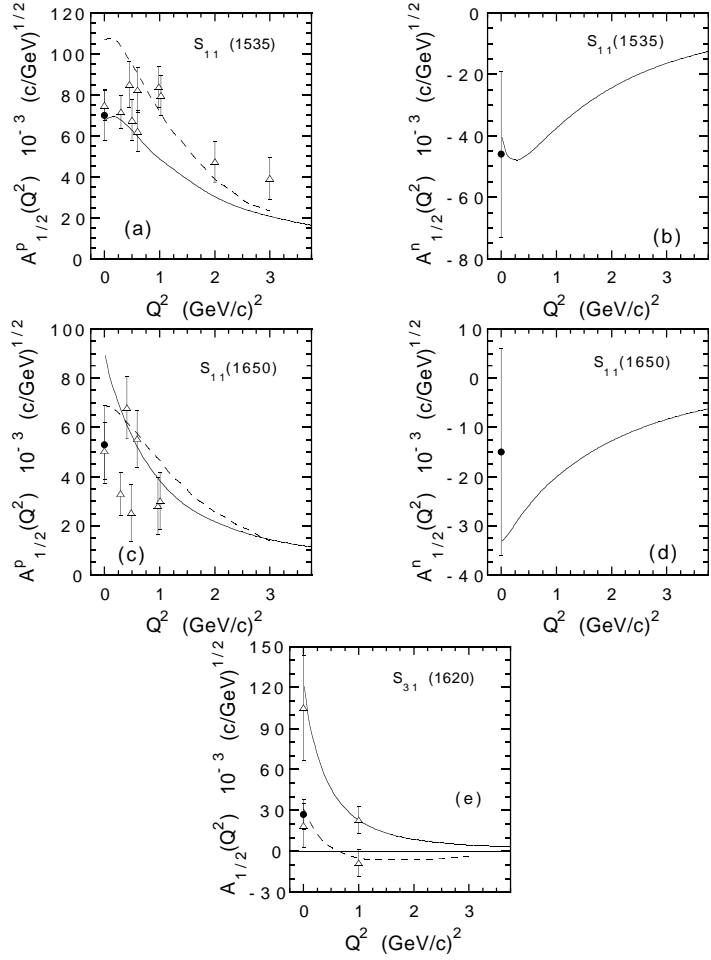} % postscript image file name
\end{center}
\vspace{-6cm}
\caption{The transverse helicities $A_{1/2}^{p(n)}$ for the nucleon 
transitions $p \rightarrow S_{11}(1535)$ (a); $n \rightarrow S_{11}(1535)$ (b);
$p \rightarrow S_{11}(1650)$ (c); $n \rightarrow S_{11}(1650)$ (d);
$p \rightarrow S_{31}(1620)$ (e), vs. $Q^2$. The non relativistic calculations
of ref. [3] are shown by the dashed lines.}
\label{tffLF}
\end{figure}

\subsection{Nucleon-Resonance transistion form factors}

Once the $CQ$ form factors have been determined
by means of the evaluation of elastic form factors within the LF model, 
one can obtain  {\em {parameter-free}} predictions for the nucleon-resonance 
transition form factors.

In Fig.~\ref{tffLF} the evaluations of the helicity amplitude
$A_{1/2}$ are shown for $N \rightarrow S_{11}(1535)$, $S_{11}(1650)$ and
$S_{31}(1620)$, and compared with the results of the non-relativistic model
of ref.\cite{aiello98}. In the case of $S_{31}(1620)$ the results for $p$ and $n$
coincide (as for $P_{33}(1232)$), since only the isovector part of the CQ
current is effective. The predictions yield an overall agreement with available
experimental data for the $P$-wave resonances and show a sizeable sensitivity to
relativistic effects, but more accurate data are needed to reliably discriminate
between different models.

\section{Electroweak structure of the nucleon}
\label{EW}

In this section I review the research activity aimed to reveal
the electroweak structure of the nucleon, in particular its strange content.
I discuss both neutrino and parity-violating polarized electron scattering.

\subsection{The strange axial current}

After the measurements of the polarized structure function of the proton,
$g_1$, in deep inelastic scattering, it  
turned out, rather surprisingly, that the constant $g^s_A$, that 
characterizes the one--nucleon matrix element of the axial strange 
current, namely
\begin{equation}
\left\langle p , s\left|\bar{q} \gamma^{\alpha} \gamma_{5} q
\right| p , s \right\rangle = 2 M s^{\alpha} g_{A}^{q}\;;\;\;\; ({\rm q=s})\;,
\label{E001}
\end{equation}
is of magnitude comparable  with the corresponding  $g^u_A$ and  
$g^d_A$  axial constants. 
(Here $p$ is the nucleon momentum, $M$ is the nucleon mass,
$s^{\alpha}$ is the spin vector and $g_{A}^{q}$ is a constant).
A theoretical analysis of deep inelastic
data led to the following values for the axial  
constants: $g^s_A = -0.10\pm 0.03$,  $g^d_A =-0.43\pm 0.03$, 
$g^u_A= 0.83\pm 0.03$ (in a more recent analysis of the
data, the value $g_A^s=-0.13\pm 0.03$ was reported).

However the values of the constants $g_{A}^{q}$ given of 
Eq.(\ref{E001}) were obtained under several assumptions. 
It is then clear that it is very important to use other methods
for the determination of the matrix elements of the strange current.
The investigation of neutral-current (NC) as well as charged-current
(CC) neutrino reactions is one of these ways \cite{alberico1,alberico2}.

\subsection{The BNL-734 experiment}

Let one consider the elastic scattering of muon neutrinos
and antineutrinos on nucleons:
$
\nu_\mu + N \to \nu_\mu + N ;\;\;
$
$
\bar\nu_\mu + N \to \bar\nu_\mu + N.
$

The cross sections of these processes depend on the electromagnetic form factors,
on the axial form factor and on the strange axial and vector form factors of the nucleon,
and the value of the constant $g_{A}^{s}$
that can be extracted strongly depends on the behaviour of the (poorly known)
axial form factor. In order to minimize this dependence 
Alberico {\it et al.}\cite{alberico1} consider the asymmetry
\begin{equation}
{\cal A}_p(Q^2) = {\displaystyle
\frac{\displaystyle
\left(\frac{d\sigma}{dQ^2}\right)_{\nu p\rightarrow \nu p} -
\left(\frac{d\sigma}{dQ^2}\right)_{{\bar\nu} p\rightarrow {\bar\nu} p} }
{\displaystyle
\left(\frac{d\sigma}{dQ^2}\right)_{\nu n\rightarrow \mu^- p} -
\left(\frac{d\sigma}{dQ^2}\right)_{{\bar\nu} p\rightarrow \mu^+ n} }
}\, 
\label{asymmdif}
\end{equation}
in order to obtain direct model independent information on
the axial ($F_A^s$) and magnetic ($G_M^s$) strange form factors 
of the nucleon. 

In ref.\cite{alberico1} the contribution of the strange form
factors of the nucleon to the NC over CC neutrino--antineutrino 
asymmetry have been calculated and compared with the information on it, which 
 one can extract from the data of the BNL--734 experiment.
In this experiment the following ratios of cross sections 
were obtained:

\begin{eqnarray}
R_\nu &=& 
\frac{\langle \sigma \rangle_{(\nu p\rightarrow \nu p)}}
{\langle \sigma \rangle_{(\nu n\rightarrow
\mu^- p)}} = 0.153 \pm 0.007 \pm 0.017
\label{rnu} \\
R_{\overline{\nu}} &=& 
\frac{\langle \sigma \rangle_{(\overline{\nu} p\rightarrow 
\overline{\nu} p)}}
{\langle \sigma \rangle_{(\overline{\nu} p\rightarrow
\mu^+ n)}} = 0.218 \pm 0.012 \pm 0.023
\label{rnubar} \\
R &=& 
\frac{\langle \sigma \rangle_{(\overline{\nu} p\rightarrow 
\overline{\nu} p)}}
{\langle \sigma \rangle_{(\nu p\rightarrow
\nu p)}} = 0.302 \pm 0.019 \pm 0.037\ ,
\label{rr}
\end{eqnarray}
where $\langle \sigma \rangle$ are cross sections folded with the experimental 
neutrino energy spectrum and integrated over the available range of momentum transfer
$Q^2$.

The neutrino--antineutrino {\it folded integral asymmetry}, 
$\langle {\cal A}_p \rangle$, is also obtained from the neutral 
current to charge current ratio of the differences between the total 
folded neutrino and antineutrino cross sections\cite{alberico2} :

\begin{equation}
\langle {\cal A}_p \rangle = \frac{
\langle \sigma \rangle_{(\nu p\rightarrow
\nu p)} - \langle \sigma \rangle_{
(\overline{\nu} p \rightarrow \overline{\nu} p)}}
{\langle \sigma \rangle_{(\nu n\rightarrow
\mu^- p)} - \langle \sigma \rangle_{
(\overline{\nu} p \rightarrow \mu^+ n)}} =
\frac{R_\nu(1-R)}{1-RR_\nu/R_{\overline{\nu}}}\, ,
\label{asymm}
\end{equation}
where the asymmetry has been written in terms of the ratios (\ref{rnu})--(\ref{rr}).
From the  experimental data one found:
\begin{equation}
\langle {\cal A}_p \rangle = 0.136 \pm 0.008 (\mathrm{stat}) 
\pm 0.019 (\mathrm{syst})\, .
\label{asymexp}
\end{equation}

\begin{figure}
\begin{center}
%\vskip -1cm
\mbox{\epsfig{file=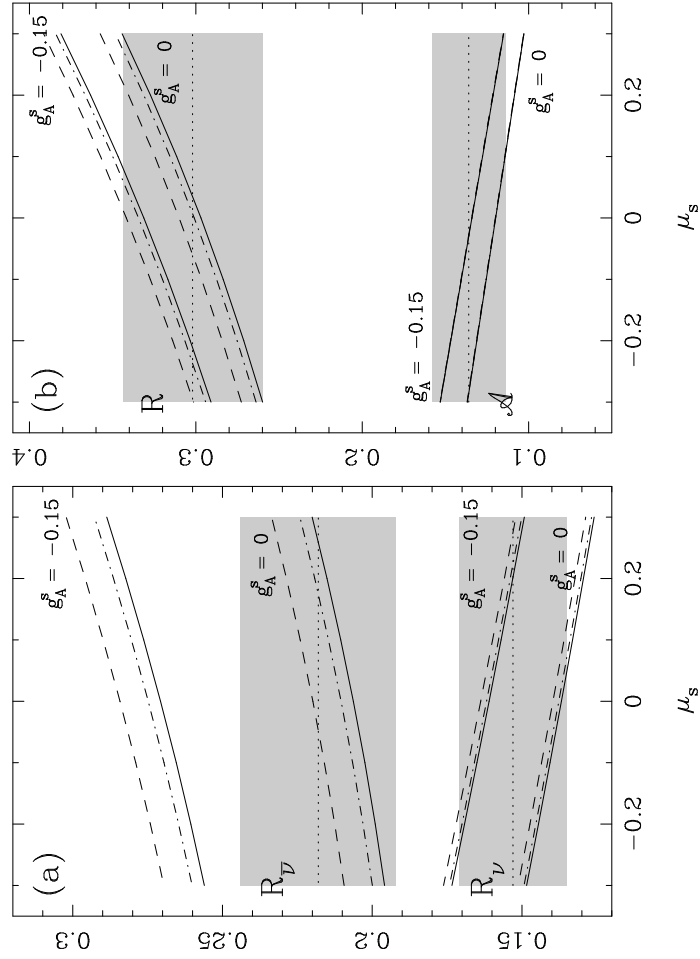,width=0.9\textwidth}}
\end{center}
\vskip -1cm
\caption{
The ratios $R_{\nu}$ and $R_{\bar\nu}$ (a), and
$R$ and $\langle {\cal A}_p\rangle$ (b), as a function of $\mu_s$:
all curves correspond to $\nu(\overline{\nu})$--p elastic scattering.
Results are shown for $g_A^s=0$ and $g_A^s=-0.15$. In both cases
we have chosen $\rho_s$ to be: $\rho_s=0$ (solid line),
$\rho_s=-2$ (dot--dashed line) and $\rho_s=+2$ (dashed line).
The shadowed regions correspond to the experimental data measured at
BNL-734 experiment.}
\label{Asymnunub}
\end{figure}

In Fig.~\ref{Asymnunub} the effects of strangeness for  
the ratios (\ref{rnu})--(\ref{rr}) and the integral asymmetry 
(\ref{asymm}) are shown. The experimental values for the various
quantities are indicated by the shadowed regions: the error band 
corresponds to one standard deviation. The usual dipole parameterization 
both for non--strange and strange form factors is assumed
$$
F_A^s(Q^2)=g_A^s G_D^A(Q^2)\;;\;\; G_M^s(Q^2)=\mu_s G_D^V(Q^2)\;;\;\;
$$
$$
G_E^s(Q^2)=\rho_s \, Q^2/4M^2\,G_D^V(Q^2)\;;\;\;\;\;{\rm with}\;\;\;
G_D^{V(A)}(Q^2)=1/(1+Q^2/M_{V(A)}^2)^{2}\;,
$$ 
while the strengths $g_A^s$, $\mu_s$ and $\rho_s$ are taken as free parameters.
The same values  for the strange cutoff masses 
as for the non--strange vector (axial) form factors are assumed.

In Fig.~\ref{Asymnunub}(a) the ratios $R_\nu$ and $R_{\overline{\nu}}$ versus
$\mu_s$ are shown for two values of the axial--strange constant: $g_A^s=0,-0.15$
and three values of the electric strange constant: $\rho_s=0,\pm 2$. 
The axial cutoff mass is $M_A=1.032$~GeV.
As it is seen from the same Fig.~\ref{Asymnunub}(a), a value 
of the strange axial constant $g_A^s$ as large as $-0.15$ is not 
favoured by the BNL--734 data.

Results of the calculation of the  ratio $R$ and the integral asymmetry 
$\langle {\cal A}_p \rangle$ are
shown in Fig.~\ref{Asymnunub}(b): 
both for $R$ and $\langle {\cal A}_p \rangle$ 
the effects induced by the axial and magnetic strange form
factors are similar. 
These effects are clearly larger (in $R$) than the ones due to the
electric strange form factor. Moreover
it is worth noticing that the integral asymmetry does not depend 
at all upon the electric strange form factor.
All the considered values of the strange parameters are compatible
with the  asymmetry  $\langle {\cal A}_p \rangle$  within the experimental
errors. However for values of $g_A^s$ as large as $-0.15$ 
the experimental value of $R$ favours $\mu_s\le 0$.

The experimental uncertainties are of the same order as the effects
of the strange form factors of the nucleon. 
Keeping this in mind and without any claim for 
a definitive evidence, the results seem to favour negative values 
of the magnetic strange parameter, $\mu_s$, if $-g_A^s$ is relatively
large.

The value of the strange magnetic form factor of 
the nucleon has been recently measured at BATES, with 
the result $G_M^s(0.1 \mathrm{GeV}^2) = 0.23 \pm 0.37 \pm 0.15 \pm 0.19$.  
This value is affected by large experimental and theoretical 
uncertainties (the last error refers to the estimate of radiative 
corrections), but it is centered around a positive $\mu_s$,
although it is still 
compatible with zero or negative values of $\mu_s$. 
Let me also notice that, if the P--odd asymmetry measured in the scattering 
of polarized electrons on nucleons 
will provide a more stringent information on the 
strange magnetic form factor, then future, precise experiments 
combining the measurement of $\nu$ and ${\bar\nu}$--proton scattering 
could allow a determination of the axial strange form factor {\it and} 
of the electric one. 

One can  conclude that the uncertainty of the available data does not allow 
to set stringent limits on the strange vector and axial--vector 
parameters, but future, more precise measurements could make 
their determination possible in a model independent way.

\subsection{Parity violating electron scattering and target asymmetry}

As already mentioned in the previous section, 
a first measurement of the PV beam asymmetry ($\Ae$) in $\vec{e}- p$ 
elastic scattering was performed at Bates/MIT 
Laboratory by the SAMPLE Collaboration giving the 
first experimental determination of the proton strangeness magnetic 
form factor at $Q^2 = 0.1 (GeV/c)^2$  
($\mus = 0.23 \pm 0.37 \pm 0.15 \pm 0.19 \mu_N$).    

Because of the difficulties inherent in the PV electron scattering 
experiment an independent determination of $\mu_s$, of the strangeness 
radius $\rdues$ and of other strangeness properties  
of the proton could be extremely useful. 

Recently Moscani {\it et al.}\cite{firenze} reported on the results of a study
on the asymmetry {\bA} of the elastic $e-{\vec p}$ scattering cross 
section (in the low $Q^2$ range) arising from the polarization
of the proton target. 
In principle, this asymmetry is even more versatile than  $\Ae$ 
for disentangling the different weak form factors because the polarization 
of the proton target can be freely chosen whereas the electron beam can 
be polarized only along the beam momentum.

The only nonzero components of the target 
asymmetry are those in the scattering plane, i.e. the transverse 
($\Ax$) and the longitudinal ($\Az$) ones (we assume the z-axis 
along the momentum transfer).

\begin{figure}
\begin{center}
\mbox{\epsfig{file=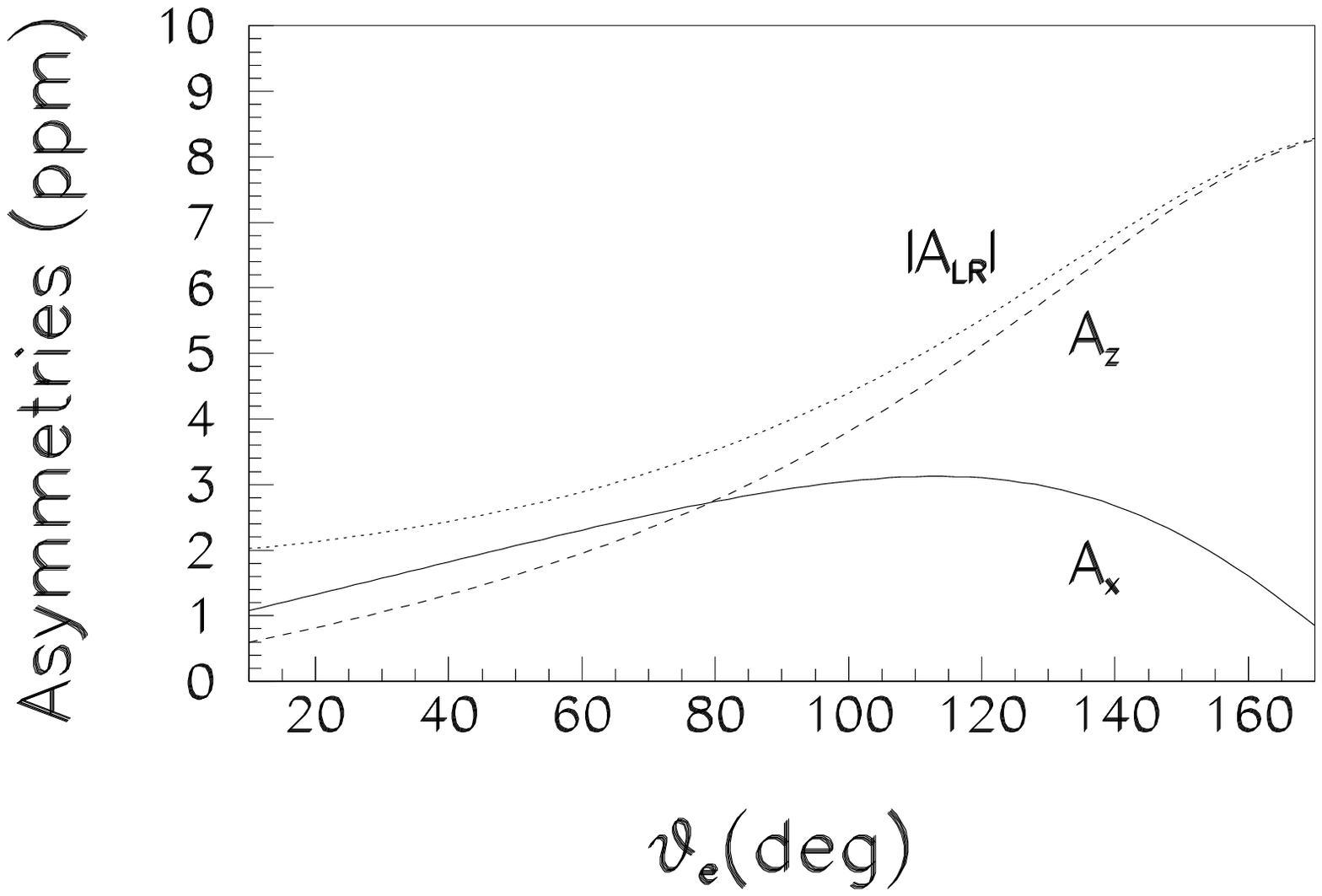,
width=0.70\linewidth,height=0.35\textheight, angle=0}}
\end{center}
\caption{Angular distribution of the target asymmetry $\Ax(\tep)$ (full line),
   $\Az(\tep)$ (dashed line) and of the modulus of the helicity asymmetry
   $\Ae(\tep)$ (dotted line) at $Q^2$=0.1$(GeV/c)^2$, with $\mus=0.23~\mu_N$
   and $\rdues=0.16$~fm$^2$.}
\label{eNAsym}
\end{figure}

In Fig.~\ref{eNAsym} the angular distribution of $\Ax, \Az$ 
and of the modulus of $\Ae$  are shown for $Q^2$=0.1 (GeV/c)$^2$, 
calculated with Jaffe's value $\rdues$=0.16~fm$^2$ and  
the central experimental value $\mus=0.23 ~\mu_N$. 
$\Ax$ and $\Az$ show a remarkably different angular dependence 
as $\tep$ increases: $\Ax \rightarrow 0$ at backward angles while $\Az$ 
reaches its maximum.   

$\Az$ does not depend on $\tGE$ (and then on $G_E^{(s)}$)  
and its dependence on $\tGA$ is lowered with respect to that 
on $\tGM$ because $\gve \ll \gae$, in particular at backward angles.  
In principle, a measurement of $\tGM$ in the target asymmetry 
is more convenient than in the helicity asymmetry making   
$\Az$ an useful quantity for a determination of $\mus$ complementary 
to that coming from $\Ae$. 
Then, a determination of $\tGA$ (and of $\gaesse$), alternative 
to that deriving from $\nu/{\bar \nu}$ scattering 
experiments could be 
carried out in this kind of PV electron scattering experiments. 

\noindent In conclusion:

\noindent i) The asymmetry $\bA$ of the elastic $e-{\vec p}$ 
scattering cross section 
arising from the polarization of the proton target 
may be a possible PV observable for an experimental determination 
of the proton weak form factors. 

\noindent ii) The most convenient decomposition of $\bA$ is obtained  
considering the proton polarization along and perpendicular to the 
momentum transfer. 

\noindent iii) The longitudinal asymmetry $\Az$ is independent of $\tGE$ allowing 
an experimental determination of the proton strangeness magnetic 
moment $\mus$.    

\noindent iv) The transverse asymmetry $\Ax$ is rather sensitive to the proton 
strangeness radius $\rdues$ in the case of backward detected electrons. 

\noindent v) A peculiarity of $\Ax$ and $\Az$ with respect 
to $\Ae$, is that their dependence on $\tGA$ 
can be enhanced over that on $\tGE, \tGM$. In fact, in the 
strict forward scattering ($\tep$ = 0$^\circ$) $\Ax$ and $\Az$ 
are determined by $\tGA$ only.

\section{Deep inelastic scattering}
\label{DIS}

Constituent quark models, on one side, and the parton picture, on the
other side, represent two complementary descriptions of the hadron
structure and the birth of QCD set the general framework to
understand deep inelastic scattering (DIS) beyond the parton model.
In the recent past a lot of work has been devoted to the attempt 
of reproducing the experimental deep inelastic structure
functions at high momentum transfer starting from a parton 
parametrization at a low resolution scale $Q_0^2$ 
where the valence contribution becomes dominant. In this way quark models, 
summarizing a great deal of hadronic properties,
may substitute low-energy parametrizations.   
Following such a path, a partonic description can be generated from gluon
radiation even off a purely valence quark system, which can be used
to generate the non perturbative input occurring in the Operator Product 
Expansion (OPE) analysis of lepton-hadron scattering in QCD.  

In the next sections I summarize some of the work done in this field by
different authors.

\subsection{Exact models}

In the parton model of DIS, it is assumed
that the final state interaction (FSI) of the struck parton with the remnants of
the target is a higher twist effect, i.e., an effect which is suppressed at
least as $m^2/Q^2$.  The qualitative motivation of this assumption is that the
time needed for the absorption of the virtual photon by the struck quark is much
smaller than the time of its hadronization and therefore in the process of
absorption the struck quark can be considered as approximately free. 

 Of course the relevance of the FSI has to be studied in the framework of
nonperturbative QCD, but in absence of a full solution it is desirable to
consider models in which the structure functions can be calculated exactly and
therefore it is possible to check whether the FSI is indeed a higher twist
effect.

Pace {\it et al.}\cite{romaDIS} investigated the role played by FSI in DIS for an
exactly solvable relativistic quark model, within the
light-cone (= front-form) hamiltonian dynamics. The confinement can be ensured by
choosing a quark-quark potential such that the mass operator of a system with a
fixed number of relativistic constituent quarks has only the discrete spectrum
(while in QCD confinement is understood as the property of the quark and gluon
Green functions to have no poles for real values of the mass).  The purpose is
to verify whether the naive treatment of confinement in relativistic CQMs is
compatible with the parton model.

The authors consider a simple system composed by two relativistic particles interacting
via the relativistic harmonic oscillator potential.  The electromagnetic
current matrix elements exhibit the correct properties under Poincar\'e
transformations and fulfill the current conservation.
In the proper Breit frame, the relevant components of the current are the same
as in the parton model.  Then, in the framework of the light-cone hamiltonian
dynamics, one can derive exact expressions for the DIS structure functions,
including, for the first time, the FSI effects calculated exactly in a relativistic
model.

\noindent Their results can be summarized in the following way:

\noindent i) the relativistic calculation differs from the nonrelativistic ones considered in
in several aspects.  In particular, the Bjorken limit implies
that one gets a finite contribution to the structure functions only from excited
states with $n\rightarrow \infty$, while the nonrelativistic approach is valid
only if $n\ll (m_0/a)^2$.  Furthermore in the infinite momentum frame only
the transverse components of the hadronic tensor survive in the Bjorken limit, 
while in the nonrelativistic case the component $W^{00}$ is the dominant one.

\noindent ii) The results could be considered as an argument in favor of the "common wisdom",
according to which the FSI in the Bjorken limit is a higher twist effect.  

\noindent  iii) The choice of the current is compatible with Poincar\'e invariance 
and current conservation.  However, these
requirements do not determine the current operator uniquely and many body
components could be present in $J^{\mu}(0)$.  Therefore one should study whether
the results of the parton model can still be recovered in the Bjorken limit if
the operator $J^{\mu}(0)$ contains many-body interaction terms.

\subsection{Covariance and DIS}

It is quite evident that relativistic effects to
the nucleon wave function, as well as, covariance requirements, are needed
even for a phenomenological description of the structure of hadrons.
To this aim I discuss a constituent quark model approach
based on a  light-front realization of the Hamiltonian dynamics and recent applied
for the calculation of both polarized and unpolarized parton 
distributions\cite{faccioli,cano-h1}. 

The parton distributions at the hadronic scale are assumed to be valence quarks 
and gluons, and their twist two component is determined by the quark momentum 
density.

\begin{figure}
\begin{center}
\mbox{\epsfig{file=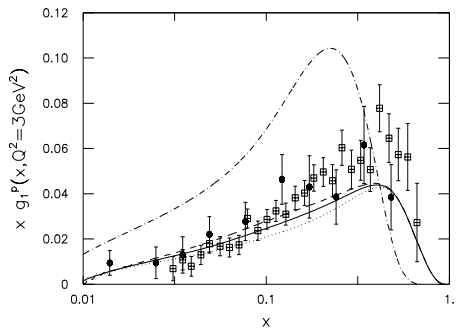,width=0.50\linewidth,height=0.25\textheight, angle=0}}
\end{center}
\vspace{3mm}
\caption{The proton polarized structure function $g_1$ at $Q^2=3$ GeV$^2$.
The full curve represents the NLO ($\overline{MS}$) results of a complete light-front 
calculation within a scenario where no gluons are present at the hadronic scale (scenario A);
the corresponding non-relativistic calculation are shown by the dot-dashed line.
A scenario including negative polarized gluons ($\int \Delta G = -0.7$)
at the hadronic scale is summarized
by the dotted line; dashed line shows the case of positive gluon 
polarization ($\int \Delta G = +0.7$).}
\label{g1LF}
\end{figure}

The relevant effects of the relativistic covariance are particularly evident looking
at the polarized distributions. In that channel the introduction of Melosh
transformations results in a substantial suppression of the responses at large 
values of $x$ and in an enhancement of the response for $x \leq 0.15$. 
The consequences can be appreciated looking at the results at the experimental scale
after a Next-to-Leading order evolution (see Fig.~\ref{g1LF}) and to the 
orbital angular momentum parton  distributions\cite{cano-oam}.
The Melosh rotation dynamics introduce the basic new  
ingredient in the calculations and its effect is quite sizeable in suppressing 
the proton response in the region $x \leq 0.4$. 

\subsection{Relativistic spin effects in Drell-Yan processes}

A complete description of the spin degrees of freedom of
quarks and antiquarks in the nucleon requires, at leading twist,
the definition of two sets of parton distributions.  
One of them, the  helicity distribution $g_1(x,Q^2)$, have been 
intensively investigated in the last few years while the so called
transversity distribution, $h_1(x,Q^2)$, has come to the attention of
theorists and experimentalists more recently in the analysis of
Drell--Yan spin asymmetries.  In fact transversity is strongly suppressed 
(by powers of $m_q/Q$) in deep inelastic lepton-nucleon scattering
and in general in any hard process that involves only one parton
distribution. In hadron-hadron collisions the chirality of the partons
that annihilate is uncorrelated and the previous restrictions do not
apply.

It is rather well known that at the hadronic scale the equality $h_1(x,Q_0^2) =
g_1(x,Q_0^2)$ is a typical outcome of non-relativistic models of the
nucleon, in which motion and spin observables are uncorrelated.  In
other words, any departure from the previous identity is a signature
of relativity in the employed hadronic model. A complete theoretical
study of $h_1$ and $g_1$ has to account for both: the relativistic
effects which distinguish $h_1$ from $g_1$ at the non-perturbative
scale, and the pQCD evolution which differs for the two structure
functions.

\vskip 8mm
\begin{tabular}{lr}
\begin{tabular}{c}
\psfig{file=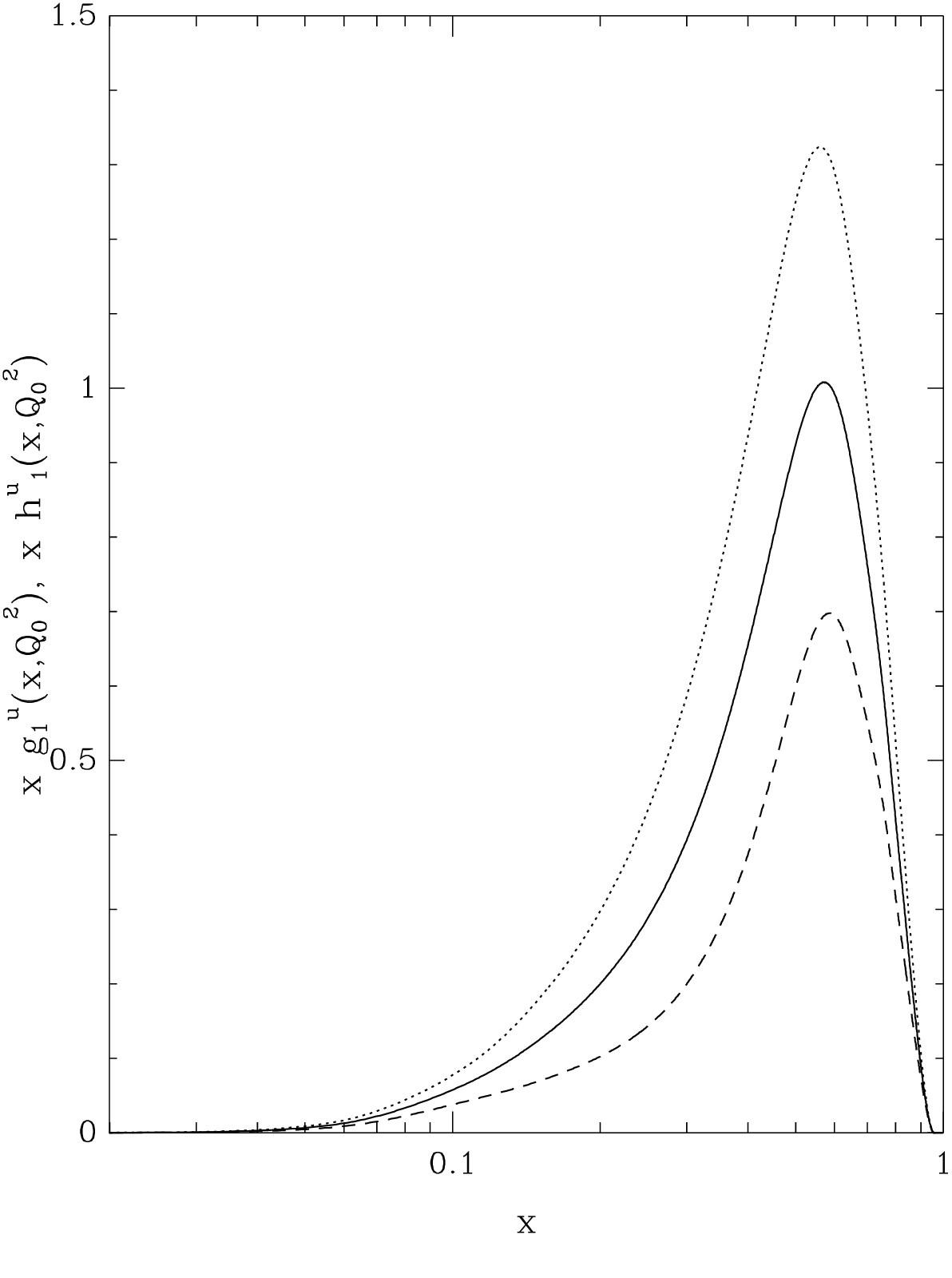,
width=0.40\textwidth,height=0.25\textheight} 
\end{tabular} & 

\begin{tabular}{c}
\rule{1.ex}{0pt} \psfig{file=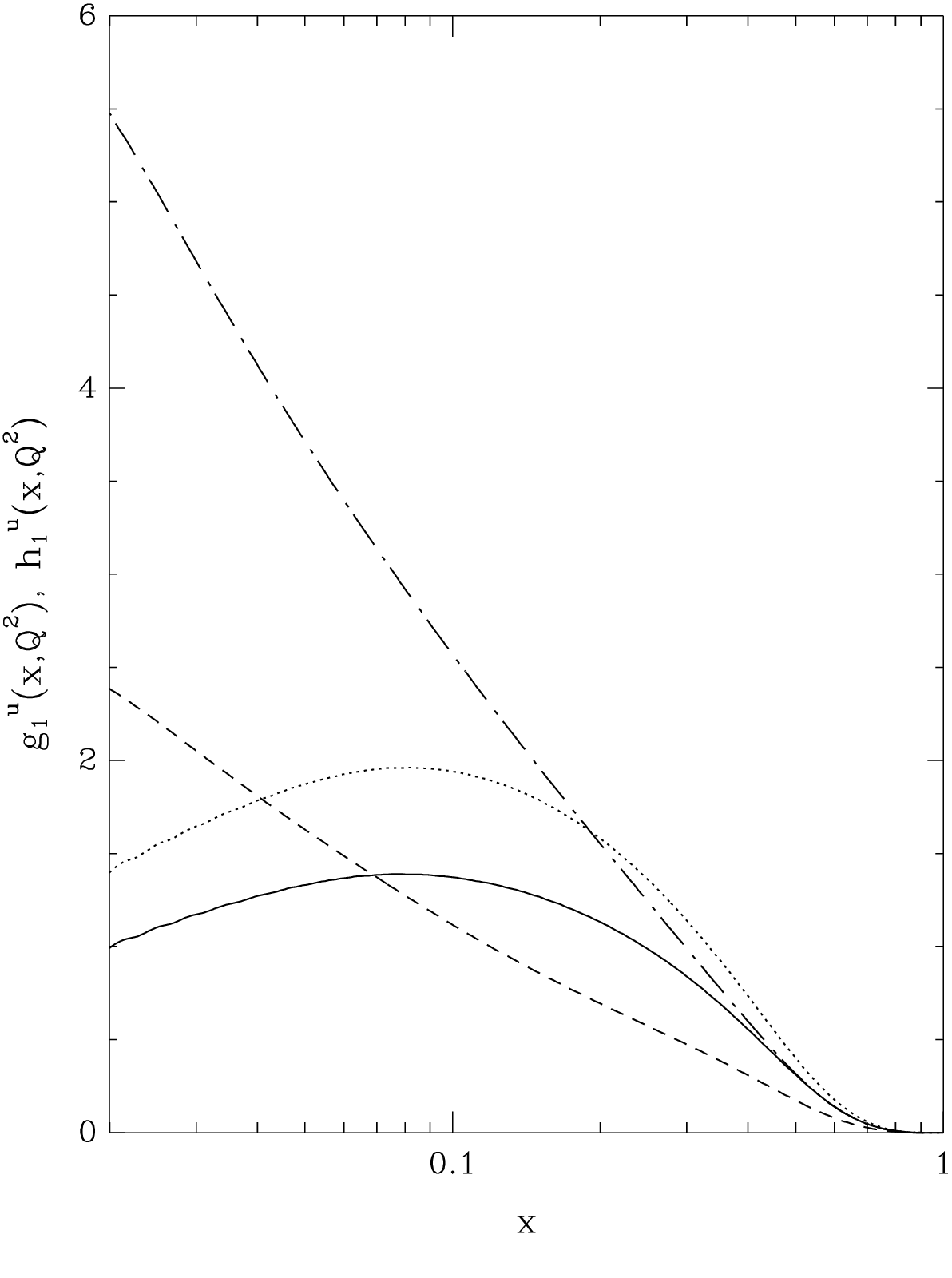,
width=0.40\textwidth,height=0.25\textheight} 
\end{tabular} 
\end{tabular}
\vskip 2mm
\noindent {Fig.~6bis.} {Helicity and transversity distributions for the
$u$ quark at the hadronic scale $Q_0^2=0.094$ GeV$^2$ (left panel) and 
after evolution up to $Q^2=100$ GeV$^2$ (right panel). 
On the left  the solid line
corresponds to $ x h_1$, the dashed line to $ x g_1$ and the dotted
line is the result when Melosh rotation is not considered
($h_1=g_1$). On the right the solid and dashed lines represent $h_1$
and $g_1$ respectively. The dotted and dash-dotted lines correspond to
$h_1$ and $g_1$ when Melosh Rotation is neglected.}
\vskip 2mm

A quantitative study of the
relativistic effects in $h_1$ and $g_1$ due to the correlations of
spin and parton motion in the hadronic systems has been 
recently completed\cite{cano-h1}. It makes use of the light-front approach previously 
discussed and the interplay between 
motion and spin is made explicit through the Melosh rotations. The
light-front covariant quark model is used
to compute the leading twist contribution to
the matrix elements at the hadronic scale $Q^2_0$. The
non-perturbative input is then evolved, at NLO, up to a higher $Q^2$
scale. 

In Fig.~6bis the results for $h_1^u$ and $g_1^u$ at the hadronic
scale $Q_0^2$ (left panel) and at the partonic scale $Q^2=100$ GeV$^2$
(right panel) are shown. A remarkable difference between $x h_1(x,Q_0^2)$ and $x
g_1(x,Q_0^2)$ appears at large $x$, reaching a peak at $x \approx
0.5$. Quantitatively they are bigger that those obtained within bag
models. It is clear that the probability of
transverse polarization is larger than the longitudinal one when
relativistic effects are considered. The results 
obtained neglecting relativistic spin-motion correlation (induced
in the light-cone approach by the Melosh rotations) are
shown also in the same figure. One gets $h_1(x,Q_0^2)=g_1(x,Q_0^2)$ as
expected. After evolution $g_1$ and $h_1$ differ mainly at low $x$ 
($x \leq 0.1$) because of pQCD evolution, while the
inclusion of the correlations between spin and motion produce large
effects also in the medium and large $x$ region.

%\vskip 8mm
\begin{figure}
\begin{center}
\mbox{\psfig{file=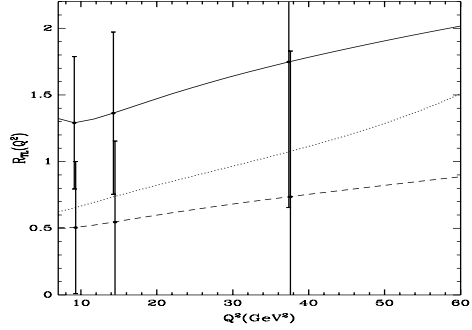,width=0.5\textwidth,height=0.25\textheight}}
\end{center}
\caption{Ratio between transverse and longitudinal parton
distributions, Eq. (\ref{ryintdef}), as a function of the invariant
mass of the produced lepton pair ($Q^2$) at a center of mass energy
corresponding to HERA--$\vec{N}$ ($\sqrt{s} = 39.2$) GeV.  Dashed line
shows results neglecting Melosh rotations, the dotted line corressponds to the
non-realtivistic model. 
Error bars have been calculated at LO and include
acceptance corrections. Error bars in the lower curve have been
slightly shifted to appreciate the overlap.}
\label{RLTLF}
\end{figure}

In order to look for relativistic spin effects, a specific 
observable has been defined in ref.\cite{cano-h1}, namely the ratio
\begin{equation}
R_{TL}(Q^2)= \frac{\int (\sum_a e_a^2 h_1^a(x_1,Q^2)
h_1^{\bar{a}}(x_2,Q^2) + (x_1 \leftrightarrow x_2)) dy } {\int (\sum_a
e_a^2 g_1^a(x_1,Q^2) g_1^{\bar{a}}(x_2,Q^2) + (x_1 \leftrightarrow
x_2)) dy } \; ,
\label{ryintdef}
\end{equation} 
where $g_1^a(x,Q^2)$ ($h_1^a(x,Q^2)$) are the lngitudinally (transverse) 
polarized parton distributions with flavor $a$ and charge $e_a$;
the arguments $x_1$ and $x_2$ are related, for Drell-Yan processes, 
to the center of mass energy $\sqrt{s}$, the invariant mass of the 
produced lepton pair $Q^2$, and the rapidity $y=\arctan(Q^3/Q^0)$:
$
x_1 = \sqrt{{Q^2}/{s}}\,e^y
$
and
$
x_2 = \sqrt{{Q^2}/{s}}\,e^{-y}
$.

The results for this ratio and for the kinematics of HERA--$\vec{N}$ are
shown in Fig.~\ref{RLTLF}  The relative insensitivity to the details of the chosen 
potential is also evident in this representation. In the error bars shown 
take into account the limited acceptance of the detectors.
While a measurement in the region $Q > 5$ GeV cannot
distinguish the importance of Melosh Rotations, in the low mass region
($Q \approx 3$ GeV) it would be possible to single out which is the
right spin-flavor basis, though some overlap between the error bars
still persists. For RHIC the acceptance corrections are too large to 
appreciate the differences.

\subsection{Semi-inclusive structure functions}

Detecting one\cite{pavia_semi} or two\cite{twohadron} of the hadrons 
produced in the high-energy scattering 
process, one is sensitive not only to the 
distribution of partons inside the target hadron, but also to 
the mechanism of hadronization, through which a quark gives rise to a jet of new 
hadrons. One is then able to measure not only distribution functions,
but also the so-called fragmentation functions.
These functions are presently considered to be very interesting and their experimental
measurement is in progress (HERMES, COMPASS, RHIC).
Neither the distribution nor the fragmentation
functions can be calculated from first principles within perturbative QCD, 
because they belong to the 
non-perturbative realm of bound states and models are required. 

\begin{figure}
\begin{center}
\mbox{\epsfig{file=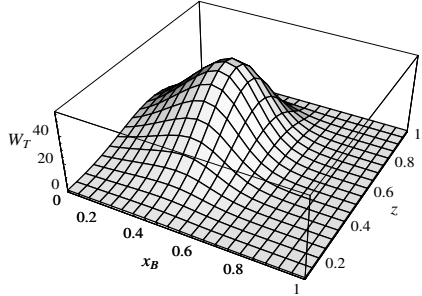,width=0.6\textwidth,height=0.30\textheight}}
\end{center}
\caption{Dependence of the structure function 
$W_T^{\,\mbox{[UU]}}$ on $x_{\scriptscriptstyle B}$ and $z$ 
at $P_{h\perp}=0$.}         
\label{fdppxz}
\end{figure}

In this context, model evaluations of the structure functions can be 
again useful. The spectator model proved to be in qualitative agreement 
with the known (transverse momentum integrated)
distribution and fragmentation functions evolved at low energies. 
Therefore, one expects it to give reasonable estimates for the convolution
integrals in semi-inclusive DIS.
A good example\cite{pavia_semi} is given 
in Fig.~\ref{fdppxz} for the reaction $e\,p\,\rightarrow e'\,\Lambda\,X$ for
unpolarized proton target and unpolarized produced $\Lambda$.

The basic assumption of the spectator model is that the 
target hadron can be divided 
into  a quark and an effective spectator state with the required quantum 
numbers, which is treated to a first approximation as being on-shell with a
definite mass. In the case of a baryon target, this second particle is a 
diquark. The quark fragments into a
jet, from which one hadron is eventually detected; the remnants of the jet are
treated effectively as an on-shell spectator state. If the detected 
hadron is a baryon, the second particle is an anti-diquark.
The vertex coupling the baryon to quark and diquark includes 
a form factor preventing the quark from being far off-shell. 
The large $p^2$-behavior of the form factor is controlled by a parameter.

An important feature of the analysis\cite{pavia_semi}
is the dependence of the cross-sections on the transverse momentum of the outgoing
hadron, $P_{h\perp}$. The measurement of this variable gives access to two new
contributions to the structure functions which have never been
observed so far, because they vanish if the cross-section is integrated over
$P_{h\perp}$. 
Furthermore, the dependence of the cross-section on $P_{h\perp}$ indirectly
tests the distribution of partonic transverse momentum inside the hadron: a
distribution largely unknown at the present.

\section{Concluding remarks}
\label{CONCL}

In the present overwiev I shortly summarized some of the italian research activities
on the field of hadronic degrees of freedom since the last two years. It seems to me 
that the contribution of the nuclear physics community to the topic contains few
specific aspects that are peculiar of a long tradition in studying complex systems.
One of them is the systematic use of many-body techniques in implementing constituent
quark models both non-relativistic and relativistic. The results are relevant for the 
interpretation of the experimental data at the TJNAF and other accelerator facilities
in the GeV region ({\it e.g.} MAMI) as illustrated by Taiuti and Pacati in their 
talks\cite{MTaiuti,FPacati}.

A second example is well illustrated by the study of the electroweak structure of the nucleon:
basic quantities like the strange content of the nucleon, can get new insights by using
the nucleus system as a filter to select interesting observables.
Addional examples can be found in the study of deep inelatic scattering. Basic questions
related to the complex dynamics of confinement and/or final state interaction can be
investigated by means of sophysticate approaches which belong to the nuclear physics knowhow.
In particular the study of the (largely) unknown non-perturbative part of
the Operator product expansion approach to deep inelastic hadronic physics, is receiving
important contributions from our community and they are relevant for the experiments at CERN,
RHIC and HERA.

\section*{Acknowledgments}

During the preparation of this contribution my wife, after a long illness, died.
I dedicate this paper to her memory asking the reader for a prayer.

\end{document}